# Quasi-periodic events in crystal plasticity and the self-organized avalanche oscillator

Stefanos Papanikolaou,[1,*] Dennis M. Dimiduk,[2] Woosong Choi,[3] James P. Sethna,[3] Michael D. Uchic,[2] Christopher F. Woodward[2] and Stefano Zapperi[4]

**When external stresses in a system – physical, social or virtual – are relieved through impulsive events, it is natural to focus on the attributes of these avalanches[1,2]. However, during the quiescent periods in between[3], stresses may be relieved through competing processes, such as slowly flowing water between earthquakes[4] or thermally activated dislocation flow[5] between plastic bursts[6,7,8]. Such unassuming, smooth responses can have dramatic effects on the avalanche properties[9]. Here we report a thorough experimental investigation of slowly compressed Ni microcrystals, covering three orders of magnitude in nominal strain-rate, that exhibits unconventional *quasi*-periodic avalanche bursts and higher critical exponents as the strain rate is decreased. Our analytic and computational study naturally extends dislocation avalanche modeling[10,11] to incorporate dislocation relaxations and reveals the emergence of the self-organized avalanche oscillator, a novel critical state exhibiting oscillatory approaches toward a depinning critical point[12]. We demonstrate that the predictions of our theory are faithfully exhibited in our experiments.**

Physical systems under slowly increasing stress may respond through abrupt events. Such jumps in observable quantities are abundant, from complex social networks to earthquakes. Even though these avalanches appear randomly sized and placed, the statistical properties of avalanches are universal, falling into well understood non-equilibrium universality classes. The main unifying concept is the *depinning* of an interface under an external field. An implicit assumption underlying these concepts is that all other coexisting physical processes are either *too fast* and thus average out, or *too slow* rendering a static approximation valid. However, the latter assumption is not always true if the slow processes rearrange the pinning landscape at rates comparable to the external field driving rates. For as the fast avalanches are scale invariant, the whole timeseries, including the waiting intervals between the fast events, is also scale invariant. It is there within the waiting intervals that a slow restructuring of the pinning field can thrive and alter universal predictions.

While intermittent plastic flow is well known[13], only recently was it shown as statistically akin to universal mean-field avalanche behavior in the quasi-static limit. Investigations of the phenomenon utilized a wide variety of techniques, including acoustic emission from deforming ice[6], high resolution extensometry of tensile strained Cu[14] and microcrystal compression tests for FCC and BCC crystals, including Ni, Al, Mo and LiF[15]. However, most of these single crystal studies covered only a narrow range of nominal high strain rates. Preliminary evidence that suggests a more complex physical picture, was discussed by some of us in Ref.[16], where a rate dependence of the cumulative strain event size distributions was observed. Interesting rate effects have also been observed in materials with solute atoms, typically polycrystalline, that display the PLC phenomenon[17, 18, 19]. The PLC avalanche distribution exponents show no evidence of strain-rate dependence (albeit strain dependence), while PLC at lower rates turns into similar-size localized slip excitations and chaotic behavior[20], distinctly different from the physical behavior observed in Ref.[16]. Instead, the PLC avalanche behavior is more consistent with the phenomenology of theories of avalanches with weakening effects[21]. In our experiments, Ni microcrystals of comparatively large dimensions, having diameters between 18 and 30 μm, were uniaxially compressed[15]. By controlling the applied external stress to maintain a *nominal* strain rate and by detecting slip with extremely sensitive extensometry, we track crystal displacements in time. In order to study the rate dependence, we perform our experiments at three different nominal strain rates ($10^{-4}$/s, $10^{-5}$/s, $10^{-6}$/s). For each sample, the timeseries of the displacement time derivative is filtered using optimal Wiener filtering methods adapted for avalanche timeseries[22], and avalanche events are appropriately defined without using thresholding.

As deformation proceeds in the micropillars, the dislocation ensemble evolves at different time scales. The most apparent activity is associated with fast glide processes which produce stochastic plastic bursts. Concurrently and in between these events, other less observable processes (Fig. 1) contribute to collective *slower relaxations*. Like glide, these too are *thermally activated processes* accessible at these high levels of stress, but have different activation barriers: for example, the viscoelastic response of the dislocation forests after fast avalanche strain bursts, the localized dislocation climb motion in directions other than the glide plane under high local stresses, and also the cross-slip processes of dislocations shifting between glide planes. They all compete to minimize the far field stress while changing the local stress landscape and bypassing the glide process. They affect dislocation slip, but at slower rate than avalanche glide[23]. In our experiments, we classify as "slow relaxation" all the deformation that does not belong to avalanches of the scaling regime. Using this definition, the slow relaxation fraction increases drastically at the two slowest strain rates. Thus, the rate dependence of the avalanche size distribution (Fig. 2(c)) occurs when the nominal strain-rate becomes comparable to the rate of the slow relaxation processes (Fig. 1). Although the exact mechanisms are unknown, one localized reorganization mechanism possible at these large local stresses and low temperatures ($0.17T_m$ (~ 300K))[23] could be tied to newly discovered unconventionally large cross-slip rates calculated for similar conditions to our experiments[24]. Regardless of the possible types of relaxation mechanisms, we focus on the experimental fact that relaxation and driving rates become comparable. We phenomenologically model the slow relaxation in an intuitive manner and then *a posteriori* show that our results are independent of the particular form of relaxation dynamics (see Supplementary Information).

The slip event sizes S, labeled by their beginning time, display a striking dependence on the driving rate.

After we smooth the timeseries over a fixed window of 400s and then rescale the time axis to display comparable strain evolution, very clear (Fig. 2(a)) oscillatory-like behavior emerges at the $10^{-6}$/s rate. The emergent period displays a strong dependence on the strain rate, while its magnitude reaches ~ 8 hrs (for $10^{-6}$/s), consistently much larger than the length we chose for the fixed window averaging (Fig. 1(c)). The novel behavior is also reflected within statistical distributions of S that show power law behavior ($P(S) \sim S^{-\tau}$) for all studied strain rates ($10^{-4}$/s, $10^{-5}$/s, $10^{-6}$/s), while the value of the power law exponent drifts from ~ 1.5 (consistent with Refs.[7, 25]) to a higher, unexpected value ~ 2.0 for the slowest strain rate (Fig. 2(c)). Analogous behavior is observed for the avalanche durations T and their correlation with the sizes.

Our explanation of the experimental data builds on the model framework of dislocations moving through a disordered landscape of forest dislocations, on a single slip plane under shear stress. This is a successful picture for avalanches during stage I plasticity[10, 11, 25] that strongly relies on well-understood models of 2+1 dimensional interface depinning[12]. We construct a minimal generalization *via* an added relaxation term, D,

$$\frac{d\phi(\mathbf{r})}{dt} = \overbrace{D\left(\frac{\sigma(\mathbf{r})}{\mu}\right)^n \Theta(\sigma(\mathbf{r}))}^{(I)} + \overbrace{\frac{1}{\mu\varepsilon}\left(\sigma(\mathbf{r}) - \sigma_f(\mathbf{r})\right)\Theta\left(\sigma(\mathbf{r}) - \sigma_f(\mathbf{r})\right)}^{(II)}, \quad (1)$$

where μ is the shear modulus of the system, and $\varepsilon \ll 1$. Here, $\phi$ is the basic slip variable of the system, the *yx*-component (Burgers vector along *x*) of the plastic distortion tensor when only infinite dislocations along *z* on *xz* slip planes are considered[10]. Part (I) of Eq. 1 denotes the coarse-grained relaxation of edge dislocations, with rate D at fixed temperature. Only positive slip motion is considered to simplify our simulations and we have shown that our conclusions are qualitatively independent of such assumptions (cf. SOM). With D we define an effective rate of thermally activated processes that lead to slow relaxation. We set the exponent *n* = 1 but our conclusions do not qualitatively depend upon it. The applied stress is the *xy*-component of the stress tensor,

$$\sigma(\mathbf{r}) = \sigma_{ext} + \sigma_{int}(\mathbf{r}) + \sigma_{hard}(\mathbf{r}) = M\,c\,t + \int d^2\mathbf{r}' K(\mathbf{r}-\mathbf{r}')\phi(\mathbf{r}') - k\phi(\mathbf{r}) \ . \qquad (2)$$

We consider a stress-controlled test in a stationary plastic regime ($\sigma_{ext} \equiv Mct$)[26], where $M$ is a machine stiffness, and $c$ has strain-rate units. The relative timescales of the relaxation and stress rate are controlled by the dimensionless parameter $R \equiv D/c$. Part (II) of Eq. 1 describes the fast glide process which drives the avalanche dynamics. Hardening is phenomenologically represented via a coefficient $k$ that controls the distance from the depinning critical point. For clarity, we separate the relevant timescales by considering $\varepsilon \ll 1$, leading to infinitely fast avalanches compared to the slow relaxations. Finally, $\sigma_{int}$ contains the appropriate interaction kernel K for single slip straight edge dislocations[10] and $\sigma_f$ denotes the uncorrelated local pinning potential due to dislocation forests. We find that our main qualitative conclusions are *independent of the kernel*, and thus are equally applicable to other models of avalanches in plasticity. In the model by Koslowski *et al.*[11] where mixed dislocations are included, one would modify our definitions by considering a single *xy* slip plane, assume $\phi$ to be the *xy* tensor component, and only apply the *zx* component of the stress.

The model of Eqs. 1 and 2 is solved by explicit integration on a two dimensional grid: For no relaxation *(D =0)*, the avalanches display statistics consistent with the predictions of the mean-field theory of interface depinning[2,8]. As $D$ increases, both the critical exponent $\tau$ for strain jump sizes $S$ $(P(S) \sim S^{-\tau})$ and the critical exponent $\alpha$ for event durations $T$ $(P(T) \sim T^{-\alpha})$, increase substantially. In the context of mean-field theory, somewhat similar behavior takes place when the driving rate $c$ is increased[22] leading to avalanche overlap and exponents decreasing below the mean-field values; we study the case where $c \to 0$ keeping $R$ fixed (and $> 1$) where the exponents increase above their mean-field values.

The increase of the exponents is accompanied by a *quasi-periodic* behavior, with intermittent but regularly spaced large slip events (Fig. 2(b)), keeping in mind that the term *quasi*-periodic is unrelated to the formal definition of quasiperiodic functions. If one considers the average avalanche size in a window, similar to the experimental study but without strain from relaxation included, it is clear that the $D = 0$ flat-in-time profile is replaced by strongly oscillating profiles in the presence of slow relaxation $(D > 0)$. The average avalanche size $(D = 0)$ is inversely related to the hardening coefficient $k$, $k \sim <S>^{-1}$. Thus, there is a distribution of hardening coefficients being effectively sampled, reflecting local heterogeneity. We assume that such a distribution $g(k')$ biases the integration, over all possible hardenings $k'$, of the size probability distribution of the $D = 0$ model, leading to the observed *dynamically integrated* size distribution. That is, a curve in Fig. 2(d) may be obtained as

$$P_{int}(S) = \int_0^\infty g(k')P(S,k')dk' , \qquad (3)$$

For the case of interest we have $P_{int}(S) = S^{-2} P(S\, k_0)$, yielding a higher effective sizes-exponent $\tilde{\tau} \equiv 2$ for slow strain rates. It is worth noting that in this picture, the largest events have a non-trivial scaling behavior (*cf.* SOM).

The profound effects of slow rate processes within our dislocation model and the comparison with experiments forces us to ask: Are our findings general? To make analytical progress toward an answer consider the *slip susceptibility* $\rho$, the multiplier giving the net number of local slips triggered by a single slip. Here, $\langle\rho\rangle$ is proportional to the hardening coefficient $(\langle\rho\rangle \propto 1-k)$. In traditional mean-field interface depinning models[1], this is the "distance" of the system from the critical depinning point and is saturated to a fixed point value after short-time transients. When $\rho << 1$, the system is far from critical, while the system is near critical when $\rho \simeq 1$. Numerical solutions to Eq. 1 verify that the additional relaxation process affects $\rho$ in an unusual way. When an avalanche with size $S_t$ takes place, $\rho$ instantaneously decreases proportionally to $S_t$, while it increases linearly in between avalanches. Minimally, we suggest that the basic physical mechanism behind the behavior of Eq. 1 (with $c \to 0$ but $R$ fixed) is given by the behavior of the slip susceptibility $\rho$, whose basic characteristics can be described by a Markov process,

$$\rho_{t+1} - \rho_t \equiv \Delta\rho_t = c_d \left(1 - \frac{S_t}{\overline{S}}\right) . \tag{4}$$

where $S_t$ is mean-field $P(S_t) = N\, S_t^{-3/2} \exp(-S_t/S_0)$ where $N$ is a normalization factor, $S_0 = a/(1-\rho_t)^2$ [1] and the step $c_d$ shall be thought as being proportional to R. The traditional avalanche mean-field behavior is described by the $c_d \to 0$ fixed point (analogous to higher experimental strain rates). The size of the avalanche at time $t$, $S_t$, is a stochastic variable which mimics the avalanche dynamics of Eq. 1. When $c_d << 1$, $\rho$ increases in small steps toward the fixed point given by $\rho_0 = 1-(\sqrt{\pi}a/2\overline{S})$ with average size $\overline{S}$ ($a$ being the minimum accessible size) (Fig. 3 left). However, there is a finite probability of a large avalanche which takes the system far from the fixed point, with $\Delta\rho_t$ large and negative. If $\delta S = S_t - \overline{S}$, then $\Delta\rho_t = -c_d \delta S/\overline{S} \sim -1$ indicates the emergence of a novel quasi-periodic behavior (Fig.1(b)) showing large negative $\rho$-jumps with $S_t$ being large *rare* avalanche events, much larger than $\overline{S}$. $\rho$ performs a *Sisyphean* task constantly ascending towards the original critical point $\rho_0$. In this way, the distribution of $\rho$ effectively flattens as $c_d$ increases (low experimental strain rates) (Fig. 3 right), leading to integrated exponents (Eq. 3). Consistently, the analogous distribution for Eq. 1 flattens as R increases (Fig. 3 right (inset)). The rare $\delta S$ events scale with $S_0 \sim 4\overline{S}^2/(\pi a)$ and qualitatively, there is a transition when $c_d \sim a\pi/(4\overline{S}) \sim 1/(S/a)$ (Fig.3 lower). We name the novel qualitative behavior as "avalanche oscillator", based on its strong resemblance to the case of relaxation limit cycle oscillations near a singular Hopf bifurcation with stochastic perturbations[27], a dynamical system similar to Sōzu, the traditional Japanese gardening device.

Contemporary observations have revealed novel and creative mechanical behaviors of single crystals in the microscale. Together with the size effects[15] and the emergence of avalanche slip events[6, 7], the importance of often neglected slow processes on intermittency has now come to light. Our experiments and theory in the world of "small" force us to reconsider our understanding of the world of "large", jammed solids and earthquake faults[28, 29, 30]. Whenever avalanches compete with other slow coexisting processes to lower stress, the nature of the dynamics shall give rise to the self-organized avalanche oscillator.

## Methods Summary

The experimental measurements were performed using the methodology described earlier[15, 16]. The data are taken at time resolutions 5, 50 and 500 Hz for different samples, depending on the case. The nominal strain rates were $10^{-4,-5,-6}$/s with corresponding average platen velocities of 4, 0.4 and 0.04nm/s, given the dimensions of the pillars. The experimental timeseries were filtered using optimal Wiener filtering methods optimized for studying avalanches[22]. In the simulations of Eq. 1, Euler time stepping is used to evolve the differential equation on a $L \times L$ grid. During an avalanche, the stress is not increased and the relaxation term (I) does not participate in the evolution. During the avalanche process, we evolve the system by using cellular automata rules: when the total local stress crosses its $\sigma_f$ threshold, the associated local slip $\phi$ increases randomly with a normal distribution with mean 1 and variance 1. The assumption of strict positivity in the local slip is used for simulation efficiency purposes, without affecting our conclusions. In the stress of Eq. 1, we have also added a term for regularizing purposes that slightly smoothens the slip profiles. It takes the form $\alpha \nabla^2 \phi$ with very small $\alpha = 0.05$. In our simulations we used a flat distribution ranged in (0,1] for the quenched disorder $\sigma_f(r)$, following a typical protocol. The kernel $K(r)$ has a continuum Fourier representation $\tilde{K}(k) = -C k_x^2 k_y^2 / (k_x^2 + k_y^2)^2$, where we set $C = 1$ for clarity purposes in our analysis. In the simulations of Eq. 4, the stochastic equation was solved using random variables that follow the required power-law distribution with exponential cutoff, generated with standard rejection methods. During solving Eq. 4 numerically, $\rho$ can jump above 1, a regime we do not consider.

**Supplementary Information** is linked to the online version of the paper at www.nature.com/nature

## Acknowledgements

We would like to thank J. Guckenheimer, C. L. Henley, E. A. Jagla, E. Nadgorny, C. O' Hern, R. Thorne, D. Trinkle, E. van der Giessen and V. Vitelli for inspiring discussions. We acknowledge support from DTRA 1-10-1-0021 (SP), DOE-BES DE-FG02-07ER-46393 (SP, WC and JPS), the US AFOSR & the AFRL, Materials and Manufacturing Directorate (DMD, CW and MU) and the ComplexityNet pilot project LOCAT (SZ).

## Contributions

DMD, MDU and CFW designed and performed the experiments. SP, DMD and CFW performed the experimental data analysis. SP, WC, JPS and SZ developed the theoretical modeling, performed the numerical simulations and carried out the data analysis. SP wrote the original manuscript and then, all authors contributed equally to improve the manuscript.

## Author Information

Reprints and permissions information is available at www.nature.com/reprints

## Affiliations

1. Department of Mechanical Engineering and Materials Science and Department of Physics, Yale University, New Haven, Connecticut, 06520-8286, USA,

2. Air Force Research Laboratory, Materials and Manufacturing Directorate, AFRL/RXLM, Wright-Patterson AFB, OH 45433, USA,

3. Laboratory of Atomic and Solid State Physics, Department of Physics, Clark Hall, Cornell University, Ithaca, NY 14853-2501, USA,

4. CNR -Consiglio Nazionale delle Ricerche, IENI, Via R. Cozzi 53, 20125 Milano and ISI Foundation,

Via Alassio 11/c, 10126 Torino, Italy.

## Competing financial interests

The authors declare no competing financial interests.

## Corresponding author

Correspondence and requests for materials should be addressed to Stefanos Papanikolaou (E-mail: stefanos.papanikolaou@yale.edu)

## Figure Legends

**Figure 1. Dislocation motion and several slow relaxation processes during avalanche waiting intervals**. **a.** Schematic toy demonstration of typical unit dislocation motions. Lighter to darker indicates time evolution. Under stress, a dislocation loop nucleates and grows until it gets pinned on its slip plane, which is a common and fast glide-slip burst unit process. Then, a screw dislocation segment undergoes *double cross-slip* to a parallel slip plane, *bypassing* glide barriers. Finally, the dislocation glides and ultimately, may climb. These unit processes are underlying to the dislocation ensemble dynamics (not shown). **b.** Deformation rate timeseries of a Ni sample at $10^{-6}$ rate. The avalanche phenomenon does not only describe fast and violent scale-invariant bursts [24], but also long waiting times [3, 38] between glide events. **c.** Estimated strain percentage accumulated in viscoplastic relaxation, with the threshold set by the event size distribution (Fig. 2)(black): The percentage (relaxation strain/total final strain) strongly increases as the rate decreases. Experimental noise contributes to the relaxation strain measured. On the right, a non-trivial *quasi-period* (see Fig. 2) of avalanche behavior emerges and increases dramatically as the nominal rate decreases.

**Figure 2. Comparison between microplasticity experiments and theoretical modeling**. **a.** Average avalanche size in 400s windows vs. time for different strain rates. Time axes are rescaled by the nominal strain rate, aligning the "strain scales". Quasi-periodic avalanche behavior emerges as the nominal strain rate decreases. The period is similar in "strain scale" – a key prediction of our theory. **b.** Stick-slip oscillations observed experimentally in **a**, form typical characteristics of the model of Eq. 1. The relaxation rate is fixed and the strain-rate is varied (by modifying c), following the experiments. The unit of strain is 2×

$10^{-6}$ and the fast timescale 0.5s. We show the *actual* avalanche events without the distortion that appears due to the strain coming from slow relaxation; this difference gives the overall scale mismatch of **a** and **b**. **c.** Drastic critical exponent increase for the sizes of the displacement jumps, as the rate decreases. **d.** As in **c**, in the model of Eq. 1 the decrease of *c* shows similar behavior with exponent drift from ~$1.5^8$, to ~2.1, with fitting error ~0.2, consistent with the discussion in the text and with fitting cutoff functional forms $f(S/S_0)$ that are discussed in the Methods.

**Figure 3. The avalanche oscillator mechanism and stochastic modeling of the slip susceptibility**. **a.** As the rate $c_d$ increases, for $\bar{S} = 0.1$, large noise causes $\rho(t)$ of Eq. 4 to oscillate between $\rho \sim 1$ and small $\rho$, causing larger exponents. **b.** The probability distribution of $\rho$. As $c_d$ increases (more diffusion, slower strain rate), any $\rho$ becomes equiprobable. In the inset, the we use $\langle S \rangle_{50}$ calculated for Eq. 1, averaged with a running window of size 50δ. The simulations of Eq. 1 have the same parameters as in Fig. 2. The histograms, shown in the appropriate scale ($1-1/\langle S \rangle \equiv \rho$ for the kernel used) shows qualitatively similar flattening behavior as Eq. 4. **c.** The novel regime with large $\rho$ fluctuations is separated from the traditional regime $\rho \approx \rho_0$. The line $c_d \sim 1/\bar{S}$ shown, as described in the text. $\tilde{\tau}$ was calculated using Eq. 3 at equidistant points with a final interpolation for the color background.

## Methods

**Experimental:** The data are taken at time resolutions 5, 50 and 500 Hz for different samples, depending on the case. The nominal strain rates were $10^{-4,-5,-6}$/s with corresponding average platen velocities of 4, 0.4 and 0.04 nm/s, given the dimensions of the pillars. Optimal Wiener filtering corresponds to a low-pass filter that has significant effects only at short timescales which are plagued by apparatus problems. In a similar fashion, as in Ref. [24], we performed adequate tests in order to confirm that the power-laws and the long-time quasi-periodic behavior are not related to the filtering procedure.

**Theoretical:** In the simulations of Eq. 1, during diffusion, Euler time stepping is used to evolve the differential equation on a $L \times L$ grid. During an avalanche, given that $\varepsilon \to 0$, the stress is not increased and the relaxation term (I) does not participate in the evolution. This approximation was performed for clarity purposes, with qualitatively similar results with the case $\varepsilon = 1$. In that case, the effect of diffusion is more visible and avalanches dissipate (for large *D*) much smaller stress percentage, since the relaxation term dominates the behavior. During the avalanche process, we evolve the system by using cellular automata rules: when the total local stress crosses its $\sigma_f$ threshold, the associated local slip $\phi$ increases randomly with a normal distribution with mean 1 and variance 1. The assumption of strict positivity in the local slip is used for simulation efficiency purposes, without affecting our conclusions as we demonstrate in the SOM[31]. In the stress of Eq. 1, we have also added a term for regularizing purposes that slightly smoothens the slip profiles. It takes the form $\alpha \nabla^2 \phi$ with very small $\alpha = 0.05$. We have checked for several system sizes (up to $64^2$) that this term does not affect our reported results in any visible manner. Also, we note that this term is physically motivated, in as much as it is connected to the coarse-grained form of the stress generated by dislocation pile-ups [8]. In our simulations we used a flat distribution ranged in (0,1] for the quenched disorder $\sigma_f(r)$, following a typical protocol. The kernel $K(r)$ has a continuum Fourier representation $\tilde{K}(k) = -C\, k_x^2\, k_y^2/(k_x^2 + k_y^2)^2$ [10], where we set $C=1$ for clarity purposes in our analysis. A modification of $C$ modifies the strength of disorder required in the model in order to observe avalanche behavior, with no other changes. In all simulations using Eq. 1 (unless explicitly mentioned otherwise) the hardening coefficient $k$ is selected from the formula $k = 2L^{0.85}/S_0$ where we chose $S_0 = 1000$ with $S_0$ being approximately equal to the cutoff size of the distribution that is derived for $D = 0$. The reason for this choice has to do with the fact that the nature of the kernel is such that a fixed local hardening coefficient $k$ does not set the cutoff for the size distribution. Rather, it allows for a weak increase with the system size. However, for our purpose (studies of $D > 0$) it was crucial to have well controlled critical distributions for $D = 0$, independent of the system size, to identify the concrete effects of the relaxation on the distributions. In all plots we refer to the value of $R = D/c$. We shall note that all our conclusions remain qualitatively unaltered if a strain-rate-controlled test is considered, while the only requirement we identified for the emergence of the avalanche oscillator is the existence of a large range of time intervals between distinct events, as self-similarity requires[32]. The independence of the avalanche oscillator behavior from the external forcing type is in contrast to typical microscopic friction stick-slip[33, 34] or coarse-grained weakening[21, 35] modeling that lead to stick-slip avalanche periodicity and typical infinite off-critical events[36]. In the simulations of Eq. 4, the stochastic equation was solved using random variables that follow the required power-law distribution with exponential cutoff, generated with the standard rejection method[37]. While solving Eq. 4 numerically, $\rho$ can jump above 1, a regime that we do not consider. There are several options to deal with the boundary condition at $\rho = 1$ which are numerically very similar for large $S$ and small $c_d$. After a jump which takes $\rho > 1$: i) $\rho$ is reset to a random value between 0 and 1, ii) $\rho$ is reset to a specific value (for example, 0 or $\rho_0$), iii) $\rho$ is returned to its previous value and the step is rejected (this method was used for the generation of Fig. 3 Center). We shall reiterate that these crossings ($\rho$

> *1*) are regularization/finite-size effects and do not define the system's behavior at long times and in the limit of $S/a \to \infty$, as we verified in both simulations of Eq. 1 (showing that the distribution "bump" consistently vanish with the system size) and Eq. 4 (showing that different treatments of the $\rho =1$ boundary lead to the same conclusion and phase boundary $c_d \sim 1/\overline{S}$ ). Finally, in Fig. 2, the fitting functional forms used were $c_0 S^{-\tau} e^{-c_1(S/S_0)^{c_3} + c_2\sqrt{S/S_0}}$ where $c_0$, $c_1$, $c_2$, $c_3$ and $\tau$ are fitting parameters. As it appears from our theoretical study, the cutoff functions $f(S/S_0)$ are rate dependent. For example, in Fig. 2(c), we find $c_3=3/2$ at $10^{-4}$ rate, while it is $c_3=1$ at $10^{-6}$. In Fig. 2(d), $c_3=2$ at the low rate, while it is $c_3=1$ at the high one.

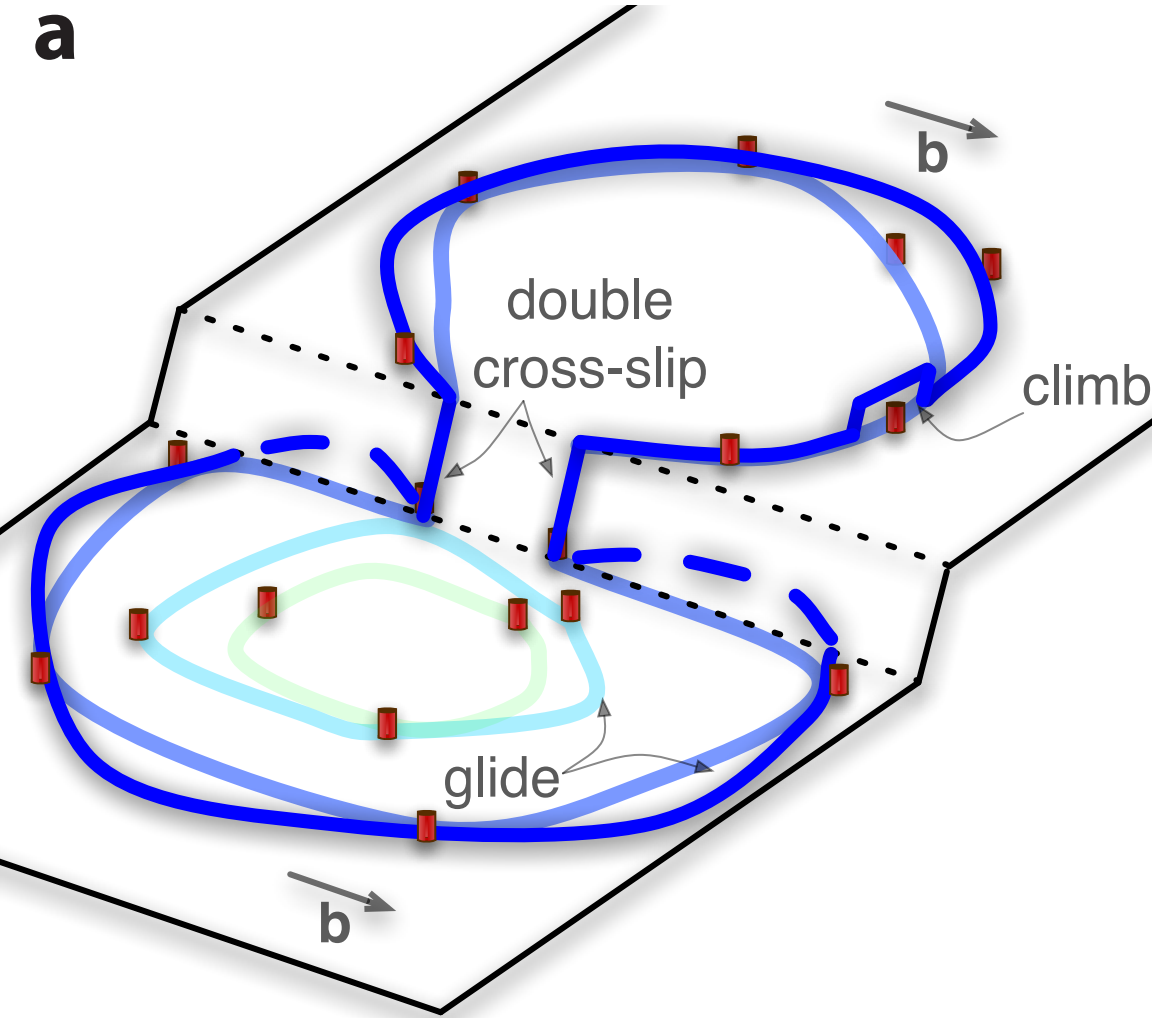 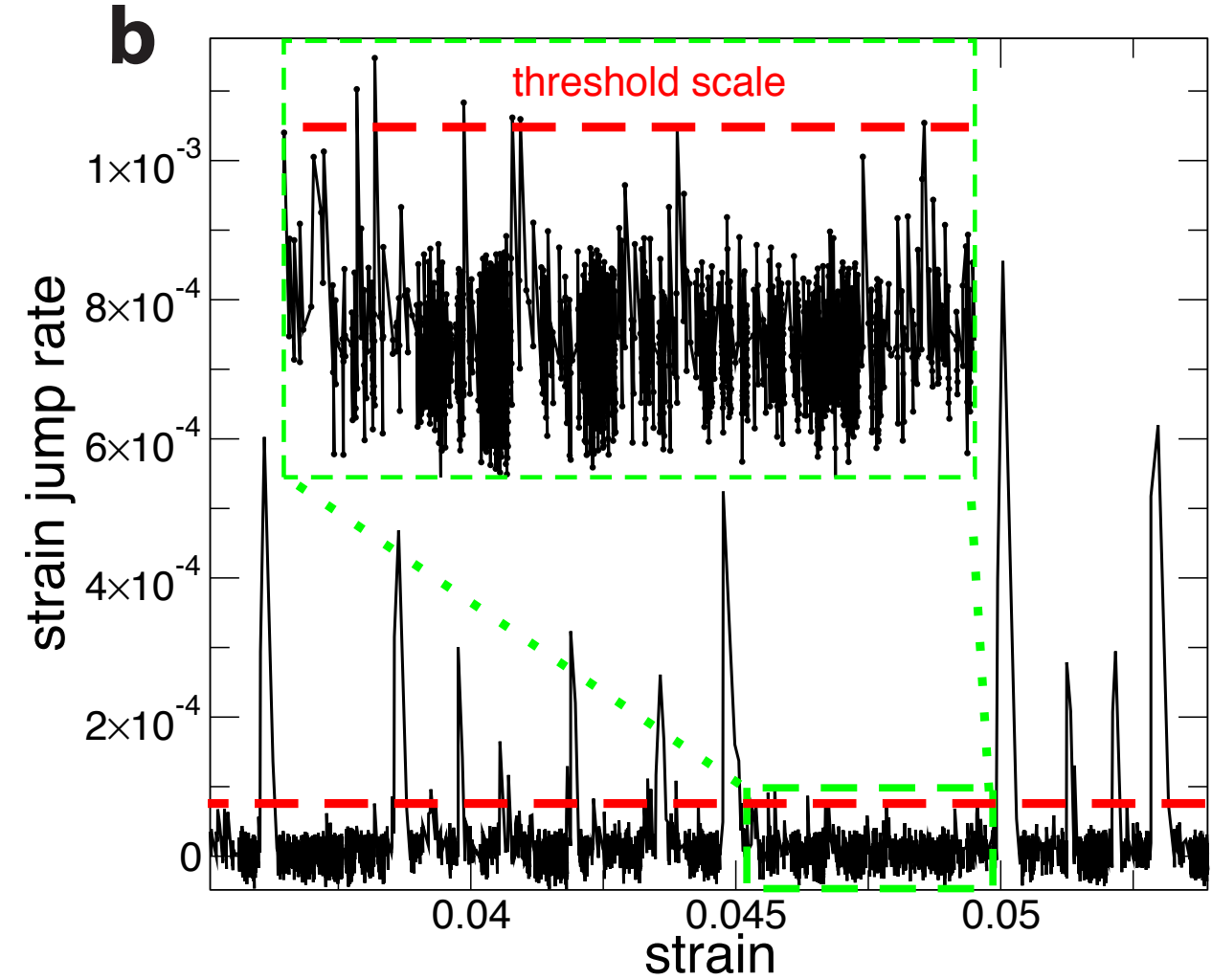 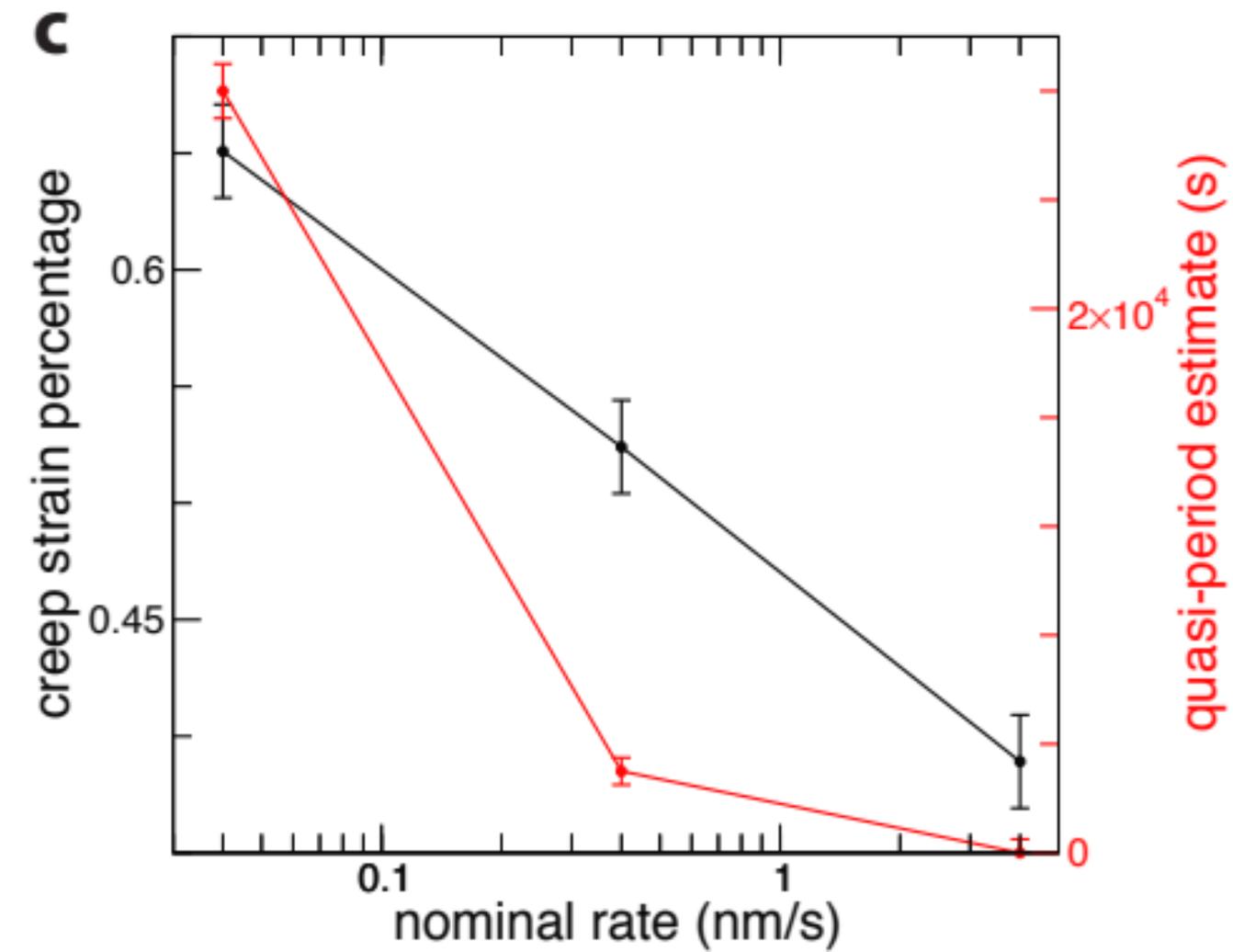

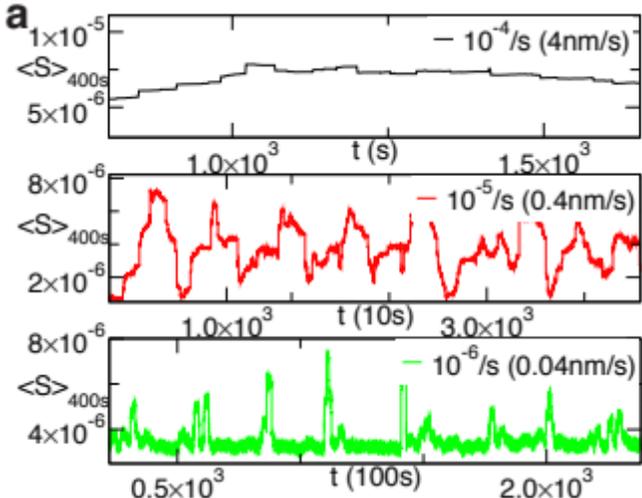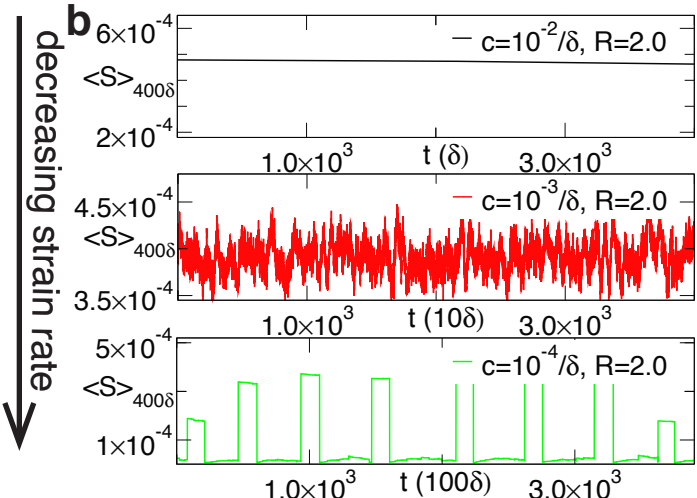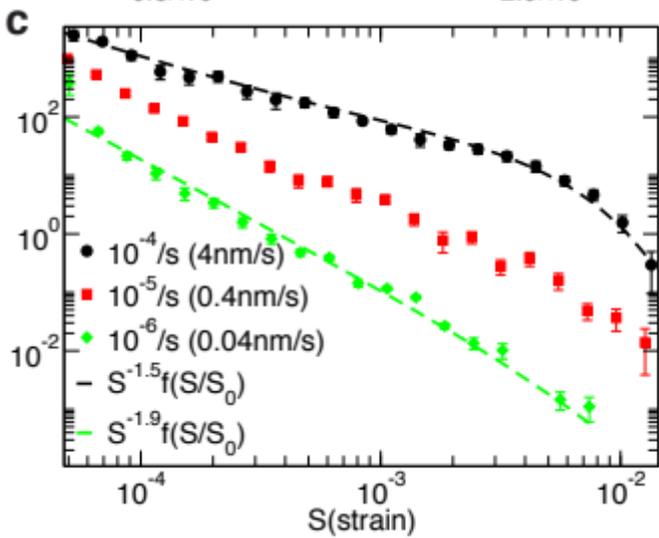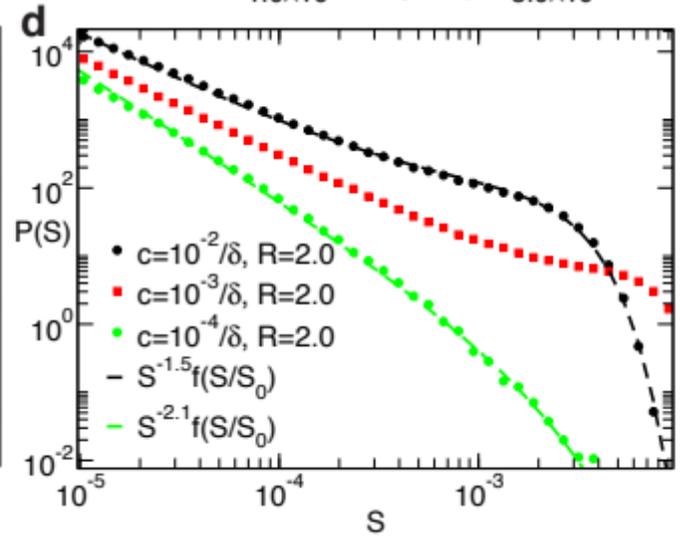

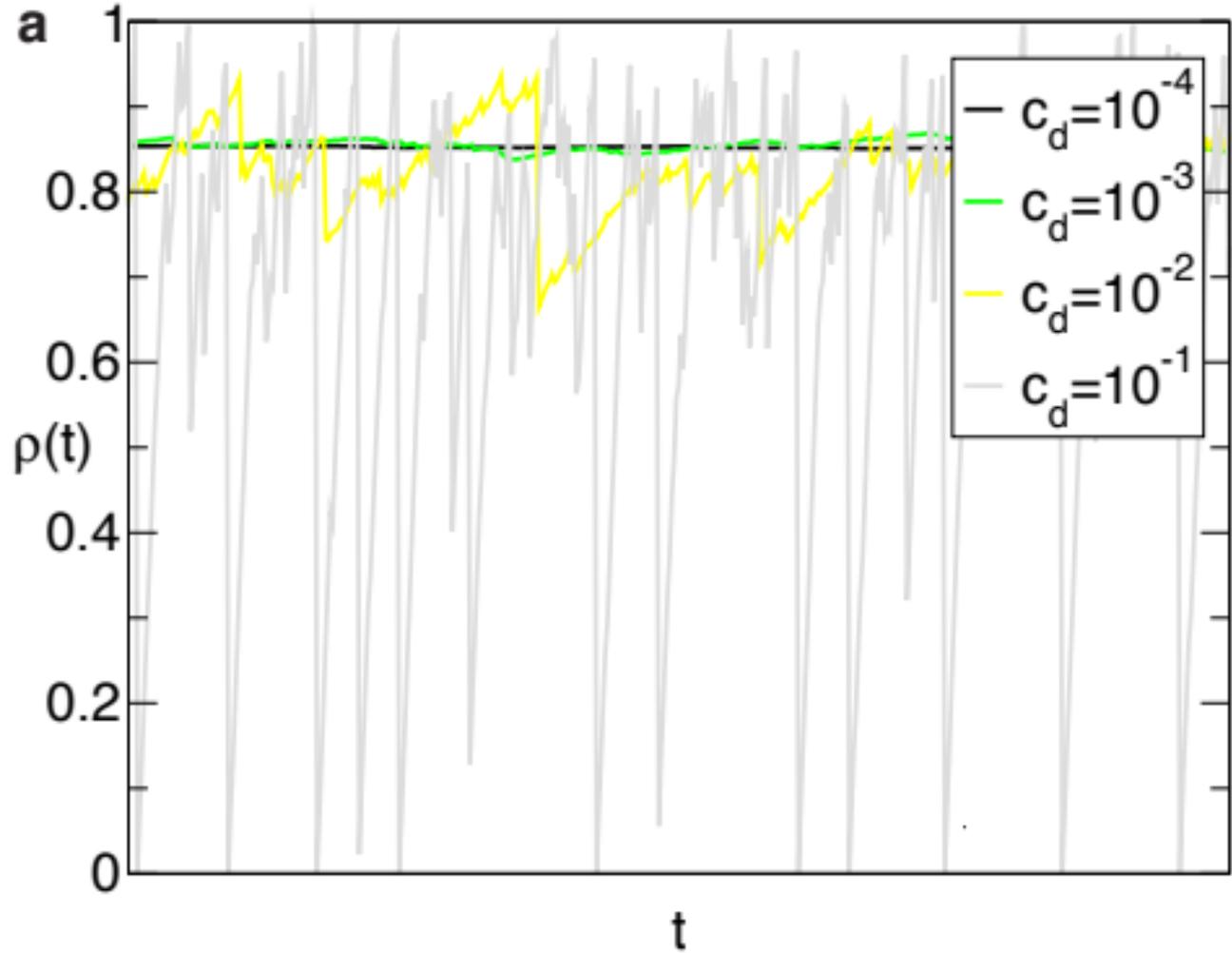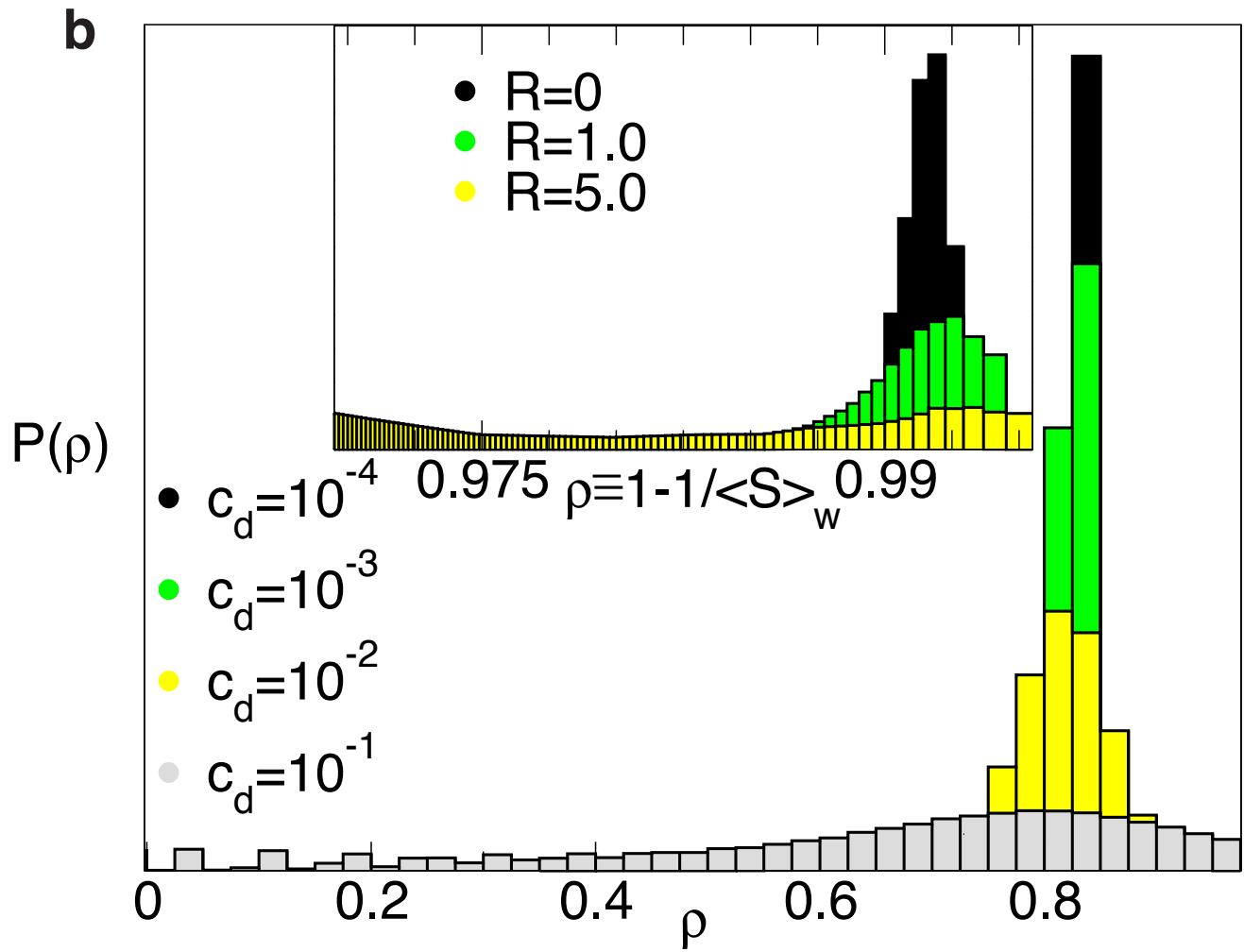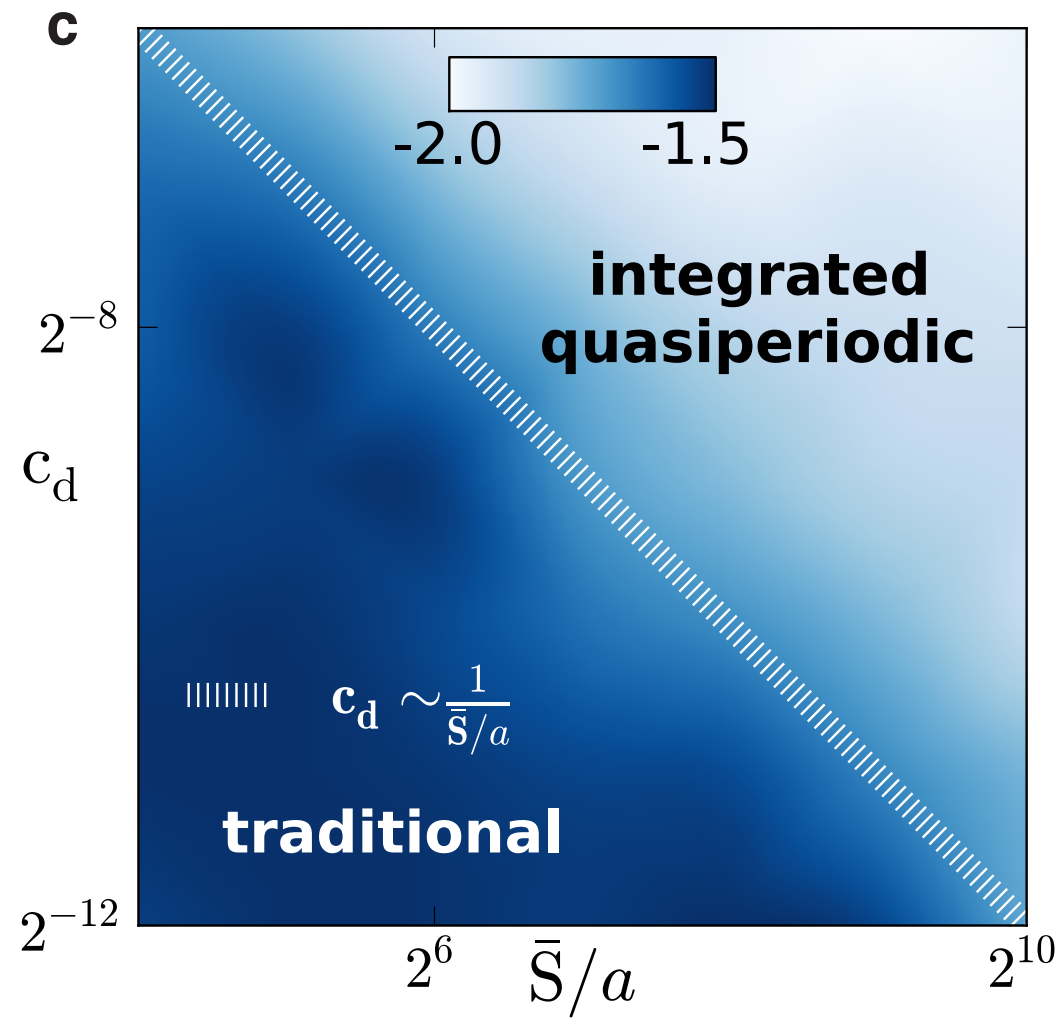